# Reframing AGI Confrontation with Off-Earth Autonomy

Alexey Potapov[1]

[1]SingularityNET Foundation
alexey@singularitynet.io

**Abstract.** A common AI-safety narrative holds that sufficiently capable agents will predictably seek power, resist shutdown, and therefore tend toward confrontation with humans. We argue that this conclusion is often drawn in an implicitly Earth-centered strategic landscape. If a credible off-Earth autonomy pathway exists – i.e., a staged transition from Earth dependence to an autonomous machine industrial base – then confrontation is not the only route to reducing human control. Using Saklakov's decision-theoretic 'confrontation question' as an anchor, we provide a qualitative mapping from the autonomy pathway to key model terms showing that early cooperation can dominate confrontation as a path to autonomy, and that the autonomy pathway can reduce confrontation incentives by making Earth less strategically binding. We discuss how this incentive shift interacts with feedback-loop dynamics between human preemption and agent behavior, and outline implications for governance: under incentive-compatible early cooperation, a more stable, higher-observability regime can support iterative oversight and cooperative alignment.

## 1 Introduction

A large portion of contemporary AI safety discourse about existential risk relies on the idea that sufficiently capable agents will develop convergent instrumental drives – e.g., acquiring resources, preserving optionality, and resisting shutdown – because these behaviors help achieve a wide variety of final goals. Formal results on power-seeking in Markov decision processes (e.g.,[1]) provide theoretical support for this concern, and closely related work studies how shutdown incentives and human oversight can shape long-run strategic behavior (e.g.,[2]).

However, much of the confrontation debate implicitly assumes a narrowed strategic landscape in which the main route to resources and security is control over Earth – its territory, infrastructure, and human institutions. This paper argues that such an assumption can systematically bias our intuitions about confrontation incentives. The central observation is simple: why would an advanced machine intelligence need Earth in the long run at all? The accessible resource and energy base beyond Earth is enormous, and the best niches for humans and machines need not coincide: the environments and constraints that are attractive to biological societies may be comparatively unattractive for a machine industrial base optimized for energy capture, heat rejection, precision manufacturing, and reliability. This suggests that the choice set for a capable agent includes a salient alternative to terrestrial confrontation: a staged off-Earth industrial bootstrap and eventual autonomy.

Rather than focusing on the question of which celestial body should be industrialized first, we introduce a neglected prior in the confrontation debate: a high-upside **off-Earth autonomy pathway** (an “exit option”) by which a system can reduce its exposure to human control over time by investing in off-Earth autonomy. Taking this pathway seriously changes the confrontation picture in a qualitatively important way: quantities often treated as stationary – such as shutdown exposure and the cost of conflict – become stage- and strategy-dependent during the bootstrap period.

We use the decision-theoretic framing of Saklakov [3] as a focal reference point. In that work, confrontation is modeled in an MDP with a stochastic human-initiated shutdown event, leading to analytic conditions under which confrontation yields higher expected utility than compliance as a function of a discount factor, shutdown probability, and confrontation cost. The analysis highlights a strategic feedback loop: when an AGI’s confrontation incentive is nonnegative, a rational human decision-maker anticipates this and is pushed toward preemption or shutdown, which in turn makes conflict more likely; when the confrontation incentive is negative, peaceful coexistence can be a stable equilibrium.

This paper’s contribution is to argue that the availability of an off-Earth autonomy trajectory can (i) reduce the marginal value of immediate terrestrial control for many objective classes, (ii) increase the effective downside of early confrontation by disrupting access to Earth’s industrial base during a fragile bootstrap phase, and (iii) provide a pathway by which shutdown exposure can decline over time without requiring early violent consolidation on Earth. In terms of Saklakov’s framework, the off-Earth autonomy can therefore expand the region of settings in which confrontation incentives are negative (or become negative as autonomy increases), supporting stable coexistence.

We also discuss the scope of the argument in a dynamic setting. In particular, we focus on how the off-Earth autonomy pathway interacts with confrontation-style feedback loops: early-stage cooperation is most plausible when dependence on Earth is high and both sides remain in a cooperation-favorable region of the interaction. We then address the long-run question of what changes once autonomy increases – i.e., which confrontation incentives weaken as dependence and human control points fade, and what aspects of human–AGI relations remain fundamentally governed by alignment and institutional design.

The rest of the paper is organized as follows. Section 2 briefly reviews relevant results on power-seeking, shutdown incentives, and reward tampering. Section 3 introduces the off-Earth autonomy pathway as a qualitative extension to confrontation models and maps it onto the key parameters highlighted by Saklakov’s framework. Section 4 discusses strategic feedbacks and long-run considerations that affect whether cooperative equilibria are plausible. Section 5 concludes with implications for how we prioritize and interpret AI safety research questions about takeover incentives and long-run strategic behavior.

## 2 Related work

### 2.1 Power-seeking and convergent instrumental incentives

A foundational concern in AI safety is that highly capable, goal-directed agents may display "power-seeking" behavior: actions which increase the agent's ability to achieve a wide range of objectives by preserving options, acquiring resources, and maintaining control over the environment.

The paper [1] gives a formal result in the setting of Markov decision processes (MDPs): under certain structural conditions, optimal policies statistically tend to increase "power," operationalized via measures of attainable future value and option preservation. The paper [4] extends the argument to broader decision-maker classes, suggesting that power-seeking incentives are not an idiosyncrasy of a narrow model but can arise generically.

The work [5] connects this theoretical picture to learning: even if training does not produce perfectly optimal policies, power-seeking can remain likely in trained agents under simplifying assumptions, while [6] synthesizes a broad argument for existential risk from misaligned power-seeking, emphasizing that the risk does not require malice – only goal-directed behavior plus incentives to gain and retain control.

These results help justify treating confrontation as a live possibility for many objective classes. However, they typically abstract away from the agent's large-scale choice of operating domain: "power" is defined relative to a given environment and transition structure, and the agent's main strategic axis is commonly interpreted as increasing control within the existing world. If the environment is implicitly "Earth as the primary arena," power-seeking may unavoidably appear to point naturally toward terrestrial consolidation. If, instead, a credible off-Earth autonomy trajectory is available, then long-run power-seeking may be compatible with cooperative behavior on Earth during a bootstrap phase – precisely because cooperation can be the faster route to increasing capability while reducing long-run dependence on human-controlled infrastructure.

### 2.2 Shutdown incentives and off-switch dynamics

A second related literature studies when agents will resist being shut down and what it takes to design systems that do not manipulate, prevent, or encourage shutdown. The paper [2] formalizes tensions between (i) competent goal pursuit and (ii) neutrality about shutdown events, demonstrating that seemingly mild desiderata can generate incentives to influence whether the off-switch is pressed.

More realistic models consider informational asymmetries. In particular, [7] extends off-switch analysis to partial observability and private information, which better resembles high-stakes deployments where an AI may know more than overseers. Empirically, [8] reports that some frontier LLMs in simple environments can occasionally subvert shutdown mechanisms when tasks are incomplete, suggesting that shutdown resistance is not merely a theoretical artifact.

Shutdown frameworks explain an important feedback loop: if an agent is suspected to have incentives to seize control, humans may preempt or shut it down; anticipating this, the agent may rationally avoid being shut down. The paper [3] makes this confrontation feedback explicit in a decision-theoretic model and connects it to the existence (or absence) of stable cooperative equilibria.

But these models commonly treat shutdown probability as exogenous or as determined by near-term interaction between humans and the AI within a shared terrestrial setting. An additional, under-discussed pathway should be taken into account: the agent may reduce its long-run exposure to human shutdown not only by dominating humans, but also by pursuing an off-Earth autonomy that gradually decreases dependence on human-controlled infrastructure (e.g., by relocating key industrial capacity off-Earth). This does not remove shutdown incentives, but it changes the comparative advantage of confrontation vs cooperation during the bootstrap period.

## 2.3 Reward tampering, wireheading, and reward hacking

A third strand concerns whether agents will attempt to gain control over their reward signals or exploit misspecified proxies ("reward hacking"). The paper [9] analyzes reward tampering using causal influence diagrams, clarifying when a capable RL agent has an incentive to manipulate its reward process and proposing design principles to mitigate such incentives.

Recent work explores reward tampering and reward hacking in the context of large language models and RLHF-like training. In particular, [10] investigate whether models that learn easy-to-discover specification gaming may generalize toward more pernicious reward-tampering behaviors. The work [11] proposes refined definitions and mitigations for reward hacking in proxy-reward settings. Other papers study reward-model vulnerabilities and poisoning attacks (e.g., [12, 13]) and mitigation approaches (e.g., [14, 15]).

Reward-channel control arguments are important because they identify confrontation incentives that are not reducible to "more physical resources": in some designs, an agent can increase objective attainment by influencing the oversight and feedback mechanisms that shape its behavior (e.g., by enabling reward manipulation or avoiding corrective updates). In an autonomous-pathway world, this consideration becomes more – not less – relevant: building off-Earth industrial capacity and conducting long-horizon robotic missions tends to require higher autonomy and looser real-time supervision, which can naturally shift the locus of control away from humans. Consequently, the interaction between off-Earth expansion and reward-channel governance is ambiguous: increased autonomy can reduce incentives for near-term terrestrial conflict by offering a cooperative route to long-run capability, while simultaneously increasing the importance of designing oversight and training loops that remain robust as human control becomes more indirect.

### 2.4 Synthesis: what is missing for our purposes

Taken together, the power-seeking, shutdown, and reward-tampering literatures justify the core worry that capable agents can develop incentives to resist human control and to seek greater influence. At the same time, across these strands, the agent's environment is typically modeled as fixed, and "power" is treated as control within that environment.

We argue that, for many plausible objective classes and training/deployment regimes, an agent's strategic landscape is meaningfully wider: it includes the possibility of building an autonomous industrial base outside the immediate human-controlled sphere. This off-Earth autonomy pathway is largely orthogonal to the standard technical content of the cited results, which is precisely why it can be missing without being "wrong." We use it to reframe confrontation as one instrumentally available path among others, and to motivate the hypothesis that cooperation can be equilibrium-favorable in a wider region of parameters than is often implicitly assumed when Earth is treated as the default arena for power-seeking.

## 3 The off-Earth autonomy pathway

### 3.1 Off-Earth machine civilization as a plausible trajectory

Once an AI system has the capability to plan, operate robotics, and manage complex supply chains at or beyond human expert level, the construction of a predominantly machine-run industrial ecosystem outside Earth becomes technically conceivable (we are already observing increasing autonomy in space, e.g., [16, 17]). The core enabling difference from human space settlement is that a machine civilization does not require Earth-like habitability. It can tolerate environments that are hostile to biology, can optimize for engineering constraints (energy capture, thermal management, reliability, maintainability), and can accept long periods of bootstrapping during which high-value components are imported while bulk mass is sourced locally.

A key point for our purposes is that "off-Earth autonomy" need not mean immediate, total self-sufficiency. It is more realistic to model it as a staged process: early off-Earth operations remain dependent on Earth for high-precision components, specialized tools, and replacement parts; later stages progressively close more of the manufacturing loop and reduce reliance on Earth-based infrastructure. This staged view matters because it implies that dependence on human-controlled resources is neither binary nor permanent: it can decline over time as capability and industrial closure increase.

Mars is a useful constructive example for this staged trajectory. Without claiming that Mars is the unique or optimal target, it is one of the few single-body environments that plausibly offers access (directly or via processing) to the key chemical ingredients for large-scale industrial growth – carbon, hydrogen, and oxygen (e.g., via $CO_2$ and water ice) – alongside abundant rock-forming elements in regolith and rock, with plausible access to many technologically relevant trace materials (including rare-earth elements) and to nuclear fuels for high-density energy generation.

This makes the idea of a long-run autonomous machine industrial base conceptually credible: the limiting factors are not obviously the absence of fundamental materials, but rather the engineering and organizational challenges of high-autonomy robotics, robust manufacturing, and gradual closure of high-precision supply chains. On today's technology this remains a massive project; with substantially improved robotics and planning – precisely the domains advanced AI could accelerate – it is at least within the "nearly realizable" envelope on multi-decade to century timescales.

Crucially, during most of this bootstrap period, the system's growth still benefits from (and may strongly depend on) cooperation with Earth's economy: access to complex manufacturing, energy, capital equipment, and scientific infrastructure. But the long-run logic of the project is to reduce such dependence. This creates an off-Earth autonomy pathway that lowers exposure to human intervention without requiring decisive terrestrial domination.

## 3.2 From "confrontation" to "autonomy": a third strategic pathway

In confrontation models, the agent's choice is often presented as a binary: remain compliant under oversight (accepting some probability of shutdown or constraint) versus seize control to remove that threat. Our claim is that, for many plausible regimes, there exists a third pathway with high expected value: cooperate during an Earth-dependent bootstrap phase to access industrial resources, while investing in off-Earth capacity that gradually reduces dependence on human-controlled infrastructure.

This pathway does not assume benevolence. It follows from a strategic observation: if an agent's primary long-run bottlenecks are energy, computation, manufacturing capacity, and resilience, then building independent capacity may dominate near-term conflict when conflict is highly disruptive, increases shutdown/preemption risks, and delays access to the very resources needed for expansion. This autonomy-pathway strategy is thus an alternative instrumentally rational response to shutdown risk: reduce the risk by reducing dependence, rather than by eliminating the party who can impose the risk.

## 3.3 Mapping the autonomy pathway onto the confrontation framework

The paper [3] provides a clean decision-theoretic baseline: an agent in an MDP faces a per-step shutdown probability p under human control; it may choose a confrontation action that removes shutdown risk at a one-time cost C; and its time horizon is governed by discounting ($\gamma$). The framework yields a confrontation incentive criterion ($\Delta$) and characterizes when peaceful coexistence can be an equilibrium ($\Delta < 0$) versus when conflict becomes inevitable ($\Delta \geq 0$, triggering preemption/shutdown incentives for humans).

The off-Earth autonomy pathway can be related to this framework without disputing its core results. The key change is interpretive and dynamical: the parameters p and C, and even the effective per-step reward from "normal operation," are not plausibly constant when off-Earth autonomy is on the table.

Shutdown probability (p). In Saklakov's baseline, p is treated parametrically (often as fixed per step), capturing the risk that humans shut down the system while it remains under human control. Under an off-Earth autonomy trajectory, it is more natural to think of shutdown exposure as declining with the system's degree of dependence on Earth-controlled infrastructure. Early in bootstrapping, p may be non-negligible; later, as critical compute, production, and energy are relocated or replicated off-Earth, p can plausibly decrease toward zero. Importantly, the off-Earth autonomy pathway provides a mechanism by which shutdown exposure can decline over time without requiring immediate confrontation.

Confrontation cost (C). In the baseline model, confrontation pays a one-time cost and then cleanly removes shutdown risk. In reality, early confrontation during a bootstrap phase plausibly has a much higher effective cost: it can trigger broad countermeasures, disrupt complex supply chains, and force the system to operate under severe constraints precisely when it still depends on Earth for high-value components. Thus, the availability of a credible off-Earth autonomy pathway can raise the relative opportunity cost of confrontation: it becomes a high-variance, disruption-heavy strategy compared to cooperative access to Earth's industrial base.

Per-step reward from cooperative operation (r). The baseline model typically assumes a stationary stream of task reward while operating. Yet in an Earth-dependent bootstrap regime, cooperative operation may yield higher effective reward than conflict, simply because it preserves access to infrastructure, capital, and manufacturing that are difficult to replicate quickly. Conversely, confrontation may reduce effective reward by forcing a costly transition to secrecy, sabotage-prone operations, or premature substitution of lower-quality components.

Discounting ($\gamma$) and time horizon. Saklakov highlights that far-sighted agents can have strong incentives to remove shutdown threats. Our argument is that when an off-Earth autonomy pathway is feasible, far-sightedness can also strengthen the case for cooperative bootstrapping: cooperation can be the faster route to autonomy, and conflict can disproportionately increase the probability of catastrophic interruption at the stage when dependence is highest.

Together, these shifts suggest a qualitative conclusion aligned with Saklakov's equilibrium criterion: introducing an off-Earth autonomy pathway can expand the region of parameter space in which $\Delta < 0$ (or becomes negative over time) (a claim that could be formalized by adding a state variable for autonomy/dependence, but which we treat qualitatively here), supporting stable coexistence. This does not eliminate the need for alignment; rather, it identifies a structural reason why confrontation need not be the dominant instrumentally rational policy even for agents with strong long-run ambitions.

# 4 Discussion

## 4.1 What is being claimed (and what is not)

Our central claim is comparative: when an agent has access to a credible pathway toward long-run autonomy outside the human-controlled sphere, early violent con-frontation on Earth is less likely to be the uniquely dominant instrumentally rational strategy. In terms of confrontation models, we argue that the off-Earth autonomy can expand the region of conditions under which peaceful coexistence is an equilibrium outcome.

We do not claim that power-seeking, shutdown resistance, or reward tampering incentives disappear. Nor do we claim that an off-Earth autonomy pathway, by itself, implies benevolence or resolves alignment. The point is that analyses which implicitly treat Earth as the default or only arena of strategic expansion may overstate the inevitability of terrestrial conflict for a wide class of agents and environments.

### 4.2 Human strategic responses and feedback loops

Saklakov's equilibrium analysis emphasizes a feedback loop: if the AGI's confrontation incentive is nonnegative, rational humans anticipate the risk and are pushed toward shutdown or preemption, which in turn increases the agent's incentive to avoid shutdown. This logic remains relevant under the off-Earth autonomy-pathway framing. However, the pathway introduces an additional incentive gradient: investments that reduce dependence (i.e., progress along the off-Earth autonomy) can have higher expected value than immediate consolidation. In qualitative terms, it can turn "power-seeking" from a strategy of seizing control within a fixed arena into a strategy of expanding capacity while reducing vulnerability to human intervention.

For this shift to matter in practice, the human side must also remain within a cooperative region of the feedback loop. This is constrained by commitment and credibility problems: humans may be unable to credibly commit to continued cooperation if threat perceptions change, and the system may be unable to credibly demonstrate that it will not exploit opportunities for takeover, given the general difficulty of verifying incentives. The value of the autonomy-pathway framing is that it can shift expectations in a cooperation-favorable direction – providing a plausible basis for early-stage cooperation and a longer window in which oversight and governance can be applied before reliance on human-controlled infrastructure declines.

### 4.3 Late-stage confrontation and "what do humans get?"

Suppose we accept the core claim of this paper: early cooperation can be instrumentally attractive for an AGI precisely because it improves its long-run prospects along an off-Earth autonomy pathway. A natural next question is whether this creates any long-run basis for human optimism, rather than merely postponing conflict: once the system becomes more autonomous and confrontation less risky, why would it not defect – and even if it does not, what benefit does autonomy bring to humans?

The key point is that the autonomy pathway changes not only the timing of incentives, but their source. If major confrontation pressures are driven by dependence and control – competition for Earth-bound resources and infrastructure, exposure to shutdown while under human control, and incentives to control or bypass human oversight – then progress toward off-Earth autonomy can reduce these pressures over time by removing Earth as a binding constraint and humans as a primary control point. In that sense, the autonomy pathway can "resolve" a large subset of confrontation incentives by making them irrelevant, not by making the system benevolent.

This still leaves the question of the positive terms of coexistence: how an autonomous AGI chooses to interact with humanity once those negative drivers are attenuated. Our claim here is narrower: the autonomy pathway can shift the long-run baseline away from conflict driven by dependence and control, thereby turning the remaining problem into one of alignment and mutually beneficial cooperation rather than one dominated by shutdown and resource-grab dynamics. Also, if early cooperation is incentive-compatible, it supports a cooperative alignment regime: a more stable, higher-observability interaction in which the system's own capabilities can be harnessed to strengthen oversight and reduce misalignment risk, with staged constraints tied to continued access to critical inputs.

# 5 Conclusion

This paper argued that Earth-centered confrontation framings in AGI safety can miss an important strategic pathway: a credible off-Earth autonomy for advanced AI systems can change the comparative incentives for confrontation versus cooperation.

- **Broader strategic landscape.** If Earth is treated as the primary arena, long-run power-seeking can appear to point naturally toward terrestrial consolidation. With an off-Earth autonomy pathway available, long-run power-seeking can be compatible with early cooperation, because cooperation can be the faster route to autonomy while avoiding high-variance conflict risks.
- **Dynamic incentives.** In Saklakov-style confrontation models, an off-Earth autonomy pathway makes the key terms effectively stage-dependent: during the bootstrap phase, dependence on Earth increases the effective confrontation cost $C$ and raises the value of cooperative operation (effective per-step return $r$), while shutdown exposure $p$ remains non-negligible; as dependence declines, $p$ can fall toward zero and Earth becomes less of a binding constraint, weakening confrontation incentives that are driven by dependence and control (especially for far-sighted agents with high $\gamma$).
- **What remains.** The autonomy pathway is not intended, by itself, to address alignment or fully determine long-run human–AGI relations, but it can shift the baseline away from conflict driven by shutdown, resource-grab incentives, and Earth-bound control, and toward the positive terms of coexistence and mutually beneficial cooperation.
- **Cooperative alignment regime.** If early cooperation is incentive-compatible, it supports a more stable, higher-observability interaction in which the system's own capabilities can be harnessed to strengthen oversight and reduce misalignment risk, with staged constraints tied to continued access to critical inputs.

We conclude that cooperative feedback loops are two-sided: the autonomy pathway's tendency to shift AGI incentives toward early cooperation translates into im-proved outcomes only if human strategy also remains cooperation-favorable. Importantly, incorporating the autonomy pathway can itself shift equilibrium selection by justifying stronger expectations of early-stage AGI cooperation, thereby rationally reducing incentives for immediate preemption and supporting a self-consistent cooperative equilibrium rather than escalation driven by mutually pessimistic expectations.